\documentclass[12pt]{article}
\usepackage{amsmath,amssymb,times,multirow,rotating,graphicx,color,fancybox}
\usepackage{times}
\usepackage{graphicx}
\usepackage{color}
\usepackage{multirow}
\usepackage[authoryear]{natbib}
\usepackage{rotating}
\usepackage{bbm}
\usepackage{latexsym}

\newcommand{\captionfonts}{\normalsize}

\newtheorem{Def}{Definition}

\makeatletter  
\long\def\@makecaption#1#2{%
  \vskip\abovecaptionskip
  \sbox\@tempboxa{{\captionfonts #1: #2}}%
  \ifdim \wd\@tempboxa >\hsize
    {\captionfonts #1: #2\par}
  \else
    \hbox to\hsize{\hfil\box\@tempboxa\hfil}%
  \fi
  \vskip\belowcaptionskip}
\makeatother   

\newcommand{\indep}{\ensuremath{{\tiny\perp\!\!\!\!\perp\!}}}
\newcommand{\E}{\ensuremath{{\mathbb{E}}}}
\renewcommand{\P}{\ensuremath{{\mathbb{P}}}}

\begin{document}
\hspace{13.9cm}1

\ \vspace{20mm}\\

{\LARGE A Distribution Free Unitary Events Method based on Delayed Coincidence Count}

\ \\
{\bf \large M\'elisande Albert$^{\displaystyle 1}$}\\
{\it Melisande.Albert@unice.fr}\\
{\bf \large Yann Bouret$^{\displaystyle 2}$}\\
{\it Yann.Bouret@unice.fr}\\
{\bf \large Magalie Fromont$^{\displaystyle 3}$}\\
{\it magalie.fromont@univ-rennes2.fr}\\
{\bf \large Patricia Reynaud-Bouret$^{\displaystyle 1}$}\\
{\it reynaudb@unice.fr}\\
{$^{\displaystyle 1}$Univ. Nice Sophia Antipolis, CNRS, LJAD, UMR 7351, 06100 Nice, France.}\\
{$^{\displaystyle 2}$Univ. Nice Sophia Antipolis, CNRS, LPMC, UMR 7336, 06100 Nice, France.}\\
{$^{\displaystyle 3}$Univ. Europ\'eenne de Bretagne, CNRS, IRMAR, UMR 6625, 35043 Rennes Cedex, France.}
%


\thispagestyle{empty}
\markboth{}{NC instructions}
\ \vspace{-0mm}\\
%
\begin{center} {\bf Abstract} \end{center}
We investigate several distribution free  dependence detection procedures, mainly based on bootstrap principles and their approximation properties. Thanks to this study, we introduce a new distribution free Unitary Events (UE) method, named Permutation UE, which consists in a multiple testing procedure based on permutation and delayed coincidence count. Each involved single test of this procedure achieves the prescribed level, so that the corresponding multiple testing procedure controls the False Discovery Rate (FDR), and this with as few assumptions as possible on the underneath distribution. Some simulations show that this method outperforms the trial-shuffling and the MTGAUE method in terms of single levels and FDR, for a comparable amount of false negatives. Application on real data is also provided.

\section{Introduction}

The eventual time dependence either between  cerebral areas or between neurons, and in particular the synchrony phenomenon,  has been vastly debated and investigated as a potential element of the neuronal code \citep{Sing1993}.  To detect such a phenomenon at the microscopic level, multielectrodes are usually used to record the nearby electrical activity. After pretreatment, the time occurrences of action potentials (spikes) for several neurons are therefore available. One of the first steps of analysis is then to understand whether and how two simultaneously recorded spike trains, corresponding to two different neurons, are dependent or not. 

Several methods have been used to detect  synchrony \citep{Perk1967,Aert1989}. Among the most popular ones, the UE method, due to Gr\"un and collaborators \citep{grunt,GDA}, has been applied in the last decade on a vast amount of real data (see for instance \citep{kilavik} and references therein). Two of its main features are at the root of its popularity: UE is not only able to give a precise location in time of the dependence periods, but also to quantify the degree of dependence by providing $p$-values for the independence tests. 

From the initial method, substantial upgrades have been developed:

(i) In the original UE method, the point processes modelling the data are binned and clipped at a rough level, so that the final data have a quite low dimension (around a few hundreds per spike train). However, it is proved in \citep{grun} that the binned coincidence count as a result of this preprocessing may induce a loss in synchrony detection of about 60\% in certain cases. The idea of \citep{grun} was therefore to keep the data at the initial resolution level despite its high dimension, but to define the notion of multiple shift (MS) coincidence count, nicely condensing the dependence feature that neurobiologists want to analyze without any loss in synchrony detection. 

(ii) The original UE method assesses $p$-values by assuming that the underlying point processes are Poisson (or Bernoulli) processes. As there is still no commonly validated and accepted model for spike trains (see for instance \cite{STAR} for a thorough data analysis), several surrogate data methods have been proposed \citep{Louis2010}. In particular, trial-shuffling methods \citep{Pipa2003,Pipaet2003} allow to assess $p$-values based on the fact that i.i.d. trials are available, using bootstrap paradigm and without making any assumption on the underlying point processes distribution. However, up to our knowledge, surrogate data methods are always based on binned coincidence count (see section \ref{coincidence} for a precise definition) whose low complexity combined to parallel programming \citep{denker} make algorithms usable in practice.

(iii) The original UE method is based on two main statistical approximations. First, it involves the underlying intensities (or firing rates) of the Poisson processes, which are unknown in practice, and so replaced by estimates. However, this plug-in procedure is not taken into account in the $p$-values computations. Then, the detection of dependence through time is done by applying several tests at once without correcting for the multiplicity of the tests. In the recent work of 
 \citep{mtgaue}, a multiple testing procedure based on a Gaussian approximation of the Unitary Events (MTGAUE) corrects those two facts, moreover including  a generalization of the notion of MS coincidence count, namely the delayed coincidence count, which does not suffer from any loss in synchrony detection. But MTGAUE is still based on the assumption that the underlying point processes are Poisson.

Our aim is here to go further by proposing a new delayed coincidence count-based multiple testing procedure, which does not need any binning preprocessing of the data as in \citep{mtgaue}, but which does not assume any model
on the underlying point processes anymore. This procedure combines a permutation approach in the line of \citep{Hoeff,romano,RW05}, with the multiple testing procedure of \citep{BH}.

To do so,  we first propose a fast algorithm to compute the delayed coincidence count, with a computational cost equivalent to the one of the binned coincidence count. 
Next we study several distribution free tests, most of them based on bootstrap approaches, as the trial-shuffling, or the permutation approach.  We finally propose a 
 complete multiple testing algorithm which satisfies similar properties as existing UE methods, but without sharing any of the previous drawbacks. Some simulations and applications to real data complete this study. 

\medskip

In all the sequel, $X^1$ and $X^2$ denote two point processes modelling the spike trains of two simultaneously recorded neurons and $X$ represents the couple $(X^1,X^2)$. The abbreviation "i.i.d." stands for independent and identically distributed. In this sense, by assuming that $n$ independent trials are observed, the observation is modeled by an i.i.d. sample of size $n$ of couples from the same distribution as $X$, meaning  $n$ i.i.d. copies $X_1,...,X_n$ of $X$. This sample is denoted in the sequel by $\mathbb{X}_n=(X_1,...,X_n)$. The corresponding probability and expectation are  respectively denoted by $\P$ and $\E$. 

Since the independence between $X^1$ and $X^2$ is the main focus of the present work, the following notation is useful:  $X^\indep$ denotes a couple $(X^{1,\indep},X^{2,\indep})$ such that $X^{1,\indep}$ (resp. $X^{2,\indep}$) has the same distribution as $X^1$ (resp. $X^2$), but 
$X^{1,\indep}$ is independent of $X^{2,\indep}$.  In particular,  the couple $X^\indep$ has the same marginals as the couple $X$. Moreover, $\mathbb{X}_n^\indep$ denotes an i.i.d. sample of size $n$ from  the distribution of $X^\indep$, and 
$\P_\indep$ and $\E_\indep$ are the corresponding probability and expectation.

Note in particular that if the two observed neurons indeed behave independently, then the observed sample $\mathbb{X}_n$ has the same distribution as $\mathbb{X}_n^\indep$.

The notation ${\bf 1}_A$ stands for a function whose value is 1 if the event $A$ holds and 0 otherwise.

Finally, for any point process $X^j$ ($j=1,2$), $dN_{X^j}$ stands for its associated point measure, defined by:
$$\int f(u)dN_{X^j}(u)=\sum_{T\in X^j} f(T) \textrm{ for all measurable function }f,$$
and for any interval $I$, $N_{X^j}(I)$ denotes  the number of points of $X^j$ observed in $I$.

\section{\label{algo}Binned and delayed coincidence counts}
\label{coincidence}
Because of the way neurons transmit  information through action potentials, it is commonly admitted that the dependence between the spike trains of two neurons is due to temporal correlations between spikes produced by both neurons. Informally, a coincidence occurs when two spikes (one from each neuron) appear with a delay less than a fixed $\delta$ (of the order of a few milliseconds). Several coincidence count functions have been defined in the neuroscience literature, and among them the classical binned coincidence count, used for instance in \citep{Pipa2003,Pipaet2003}.

\begin{Def}
The binned coincidence count between point processes $X^1$ and $X^2$ on the interval $[a,b]$ with $b-a= M\delta$ for an integer $M\geq 2$ and a fixed delay $\delta>0$ is given by
$$\psi_\delta^{coinc}(X^1,X^2)=\sum_{\ell=1}^M {\bf 1}_{N_{X^1}(I_\ell)\geq 1}{\bf 1}_{N_{X^2}(I_\ell)\geq 1},$$
where $I_\ell$ is the $\ell$th bin of length $\delta$, i.e. $[a+(\ell-1)\delta,a+\ell\delta)$ and 
$${\bf 1}_{N_{X^j}(I_\ell)\geq 1}=\left\{\begin{array}{ll}1 &\textrm{if there is at least one  point of }X^j\mbox{ in the }\ell\mbox{th bin},\\
0 &\textrm{if there is no point of }X^j\mbox{ in the }\ell\mbox{th bin.}\end{array}\right.$$
\end{Def}
More informally, the binned coincidence count is the number of bins that contain at least one spike of each spike trains.
The binned coincidence count computation algorithm is usually performed on already binned data. Therefore, given two sequences of $0$ and $1$ of length $(b-a)\delta^{-1}$,  $2(b-a)\delta^{-1}$ operations are needed to compute the binned coincidence count, without counting the number of operations needed for the binning preprocessing.

The more recent notion of delayed coincidence count, introduced in \citep{mtgaue}, is a generalization of the multiple-shift coincidence count, defined in \citep{grun} for discretized point processes, to non necessarily discretized point processes.

\begin{Def}
The delayed coincidence count between point processes $X^1$ and $X^2$ on the interval $[a,b]$ is given by
$$\varphi_\delta^{coinc}(X^1,X^2)=\int_a^b\int_a^b {\bf 1}_{|u-v|\leq \delta} dN_{X^1}(u) dN_{X^2}(v),$$
\end{Def}
More informally, $\varphi_\delta^{coinc}(X^1,X^2)$ is the number of couples of spikes (one spike from $X_1$ and one from $X_2$) appearing in  $[a,b]$ with delay at most equal to $\delta$.

The delayed coincidence count $c:=\varphi_\delta^{coinc}(X^1,X^2)$ can be computed using the following algorithm.

\begin{center}
\Ovalbox{
\begin{minipage}{14cm}
\begin{center}
{\bf Delayed coincidence count algorithm} 
\end{center}

 Given two sequences $x_1$ and $x_2$ of ordered points with respective lengths $n_1=N_{X^1}([a,b])$ and $n_2=N_{X^2}([a,b])$, representing the observations of two point processes $X^1$ and $X^2$,  

\begin{tabular}{l}
- Initialize $j=1$ and $c=0$.\\
- For $i=1,...,n_1$, \\
\begin{tabular}{ll}
&1. Assign $x_{low}= x_1[i]-\delta$.\\
&2. While $j\leq n_2$ and $x_2[j]<x_{low}$, $j=j+1$.\\
&3. If $j>n_2$, stop. \\
&4. Else (here necessarily, $x_2[j]\geq x_{low}$), \\
&\quad 4.a Assign $x_{up}=x_1[i]+\delta$ and $k=j$. \\
&\quad 4.b While $k\leq n_2$ and $x_2[k]\leq x_{up}$, $c=c+1$ and $k=k+1$.\\
\end{tabular}
\end{tabular}
\end{minipage}
}
\end{center}

It is easy to see that the complexity of this algorithm is not governed only by the lengths $n_1$ and $n_2$ but also by the random numbers of points in  intervals of length $2\delta$ for step 4.b. More precisely, it is bounded by $3n_1$ (for steps 1, 3 and 4.a), $n_2$ (for all steps 2 on all points of $x_1$) and $2n_1$ times the number of points of $x_2$ in a segment (namely $[x_{low},x_{up}]$) of length $2\delta$ (for step 4.b). In average, if $X^1$ and $X^2$ are for instance independent homogeneous Poisson processes of respective intensities $\lambda_1$ and $\lambda_2$, at most $3\lambda_1(b-a)+\lambda_2(b-a)+2\lambda_1(b-a)(2\delta\lambda_2)$ operations are required. So, for typical parameters ($(b-a)=0.1$s, $\delta=0.005$s, $\lambda_1=\lambda_2=50$Hz),  $40$ operations in average are required to compute the binned coincidence count, against $25$ operations for the delayed coincidence count. Both algorithms are therefore linear in $(b-a)$ with a slight advantage for the delayed coincidence count which exploits the sparsity of the spike trains, in the usual range of parameters. Notice that all surrogate data methods (see \citep{Louis2010}) could in principle be applied on this new  notion of coincidence, at least when only two neurons are involved.

\section{Some distribution free independence tests}
Given the observation of a $n$ sample $\mathbb{X}_n=(X_1,\ldots,X_n)$ corresponding to $n$ different trials, the aim is to test:
$$(H_0)\textrm{ "$X^1$ and $X^2$ are independent on $[a,b]$" }$$
 against 
 $$(H_1)\textrm{  "$X^1$ and $X^2$ are not independent on $[a,b]$"}.$$
All existing UE methods are based on a  statistic equal to the total number of coincidences:
$${\bf C}={\bf C}(\mathbb{X}_n)=\sum_{i=1}^n \varphi(X_i^1,X_i^2),$$
where $\varphi$ generically denotes   either $\varphi_\delta^{coinc}$, or $\psi_\delta^{coinc}$, or other coincidence count functions that practitioners would like to use (see \citep{bootnous} for other choices).

To underline what is observable or not, when ${\bf C}$ is computed on the observation of $\mathbb{X}_n$, it is denoted by ${\bf C}^{obs}$, the total number of observed coincidences. 

In the following,  several of these UE methods or testing procedures are described, which all rely on the same paradigm: "reject $(H_0)$ when ${\bf C}^{obs}$ is significantly different from what is expected under $(H_0)$". More precisely,  the independence $(H_0)$ is rejected and the dependence is detected when a quantity, based on the difference between the observed coincidence count and what is expected under $(H_0)$, is smaller or larger than some critical values. Those critical values are obtained in various ways, each of them being peculiar to each method. Note that the following procedures could be applied to any chosen coincidence count function $\varphi$, though the implemented procedures of the present simulations and data analysis only use the delayed coincidence count, that is $\varphi=\varphi_\delta^{coinc}$.

\subsection{\label{naiv}A  naive approach and the centering issue} 

As noticed above, the only question, when considering UE methods, is how to construct the critical values.

If the values of the expectation and the variance of ${\bf C}$ under $(H_0)$, that is 
$$c_0=\E_\indep( {\bf C}) \mbox{ and }v_0=\E_\indep\left(\left({\bf C}-c_0 \right)^2\right),$$ 
are precisely known, then the classical Central Limit Theorem gives  under independence that
\begin{equation*}
\frac{{\bf C}(\mathbb{X}_n^\indep)-c_0}{\sqrt{v_0}} \overset{\mathcal{L}}{\underset{n\to\infty}{\longrightarrow}} \mathcal{N}(0,1).
\end{equation*}
Then, given $\alpha$ in $(0,1)$, the test which consists in rejecting $(H_0)$ when $({\bf C}^{obs}-c_0)/\sqrt{v_0}$ is larger than $z_{1-\alpha}$, the $1-\alpha$ quantile of a standard Gaussian distribution, is  asymptotically of level $\alpha$. It means that, for this test, the probability of rejecting independence, whereas  independence holds, is asymptotically (in $n$, the number of trials) smaller than the prescribed $\alpha$. 

In the present point processes framework, strong distribution assumptions, for which the values of $c_0$ and $v_0$ are precisely known, are unrealistic. Even if the spike trains are assumed to be homogeneous Poisson processes as in \citep{mtgaue}, those quantities depend, through some formulas, on the unknown firing rates that have to be estimated and plugged into these precise formulas. It has been shown that this modifies the asymptotic variance shape, and tests under Poisson assumptions with unknown firing rates  have been developed in \citep{mtgaue}.

Since Poisson assumptions are quite doubtful on real data \citep{STAR}, the aim of the present work is to go further by not assuming any Poisson or other model assumptions for the spike trains. In this sense, the aim is to develop "distribution free" methods that are completely agnostic with respect to the underlying distribution of the spike trains.  In this case, a preliminary  step is to estimate $c_0$, only using the sample $\mathbb{X}_n$ without any distribution assumption.
Note that 
$$c_0=\E_{\indep}\left[\sum_{i=1}^n \varphi(X_i^{1,\indep},X_i^{2,\indep})\right]=n \E_{\indep}\left[\varphi(X^{1,\indep},X^{2,\indep})\right],$$
and that for $i\not = i'$, as $X_i$ is always assumed to be independent of $X_{i'}$, 
\begin{equation}\label{indepegal}
\E\left[\varphi(X_i^1,X_{i'}^2)\right]= \E_{\indep}\left[\varphi(X^{1,\indep},X^{2,\indep})\right].
\end{equation}
Therefore, $c_0$ can always be estimated in a distribution free manner by
$$\hat{\bf C}_0(\mathbb{X}_n)=\frac{1}{n-1}\sum_{i\not = i'} \varphi(X_i^1,X_{i'}^2).$$
Hence a reasonable test statistic would be based on the difference:
$${\bf U}={\bf U}(\mathbb{X}_n) = {\bf C}(\mathbb{X}_n)-\hat{\bf C}_0(\mathbb{X}_n),$$
its observed version being denoted by ${\bf U}^{obs}$.
Here, ${\bf U}(\mathbb{X}_n)$ is not an empirical mean, but a $U$-statistic, so  it does not satisfy the classical Central Limit Theorem. Hence, its limit distribution under $(H_0)$  is not as straightforward as usual.  Nevertheless, some asymptotic theorems, proved in \citep{bootnous}, show that under mild conditions (always satisfied in practice in the present cases)  the following convergence result holds:
\begin{equation}
\label{ConvUstat}
 {\bf Z}(\mathbb{X}_n^\indep)=\frac{{\bf U}(\mathbb{X}_n^\indep)}{\sqrt{n} \hat{\sigma}(\mathbb{X}_n^\indep)} \overset{\mathcal{L}}{\underset{n\to\infty}{\longrightarrow}} \mathcal{N}(0,1),
 \end{equation}
where
$$\hat{\sigma}^2(\mathbb{X}_n) = \frac{4}{n(n-1)(n-2)} \sum_{i,j,k \mbox{ all different}} h(X_i,X_j) h(X_i,X_k),$$
with
$$h(x,y)= \frac{1}{2} \Big[\varphi(x^1,x^2)+\varphi(y^1,y^2)-\varphi(x^1,y^2)-\varphi(y^1,x^2)\Big].$$
As above, denoting by ${\bf Z}^{obs}$ the quantity ${\bf Z}$ computed on the observed sample, 
 \eqref{ConvUstat}  implies that for some fixed $\alpha$ in $(0,1)$, the test that consists in rejecting $(H_0)$ when  ${\bf Z}^{obs}\geq z_{1-\alpha}$,   is asymptotically of level $\alpha$.

\begin{figure}[h!]
\hspace{-4cm}
\includegraphics[scale=0.8]{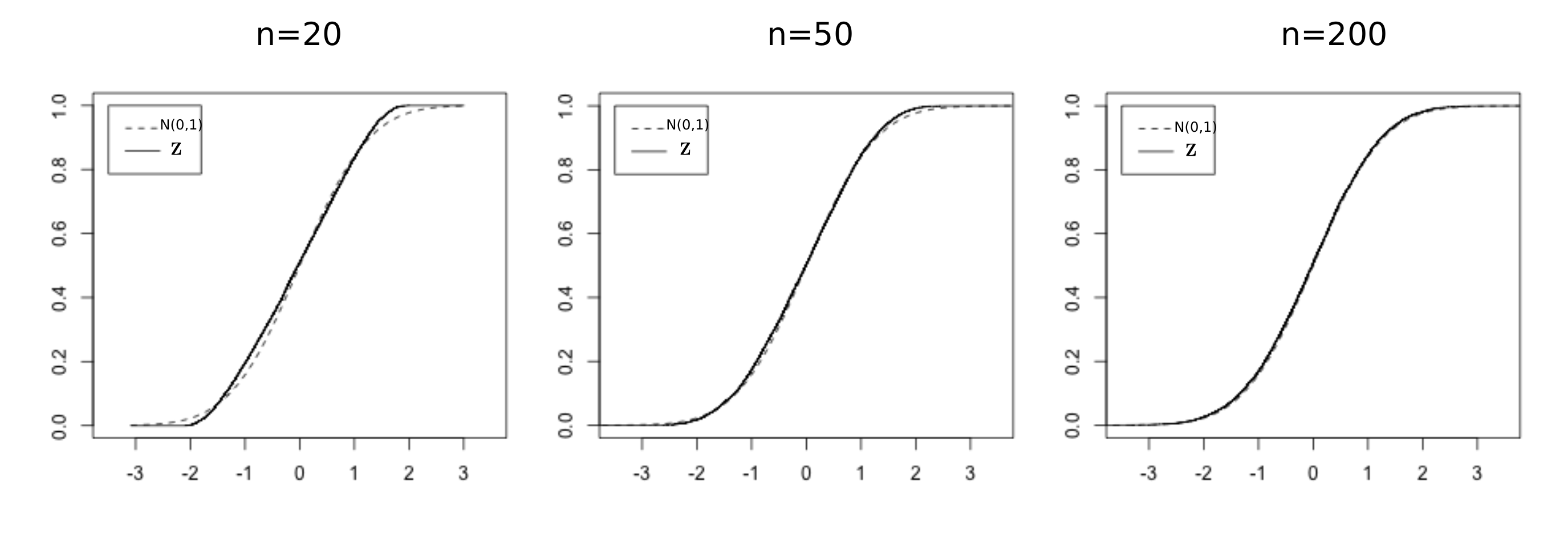}
\caption{\label{Z} Gaussian approximation of the distribution of ${\bf Z}$. In plain black, cumulative distribution function (c.d.f.) of ${\bf Z}$ under $(H_0)$, that is of ${\bf Z}^{\indep}={\bf Z}(\mathbb{X}_n^\indep)$ obtained with $2000$ simulations of $\mathbb{X}_n^\indep$, for $n=20$, $50$ or $200$ trials of two independent Poisson processes of firing rate $30$Hz, on a window of length $0.1$s with $\delta=0.01$s.  The dashed line corresponds to the standard Gaussian c.d.f.}
\end{figure}

The approximation properties of \eqref{ConvUstat} are illustrated on Figure \ref{Z}.

Clearly, one can see that the distribution approximation is good when $n$ is large ($n=200$) as expected, but not so convincing for small values of $n$ ($n=20$, or even $n=50$), particularly in the tail parts of the distributions. 
However, as it is especially the tails of the distributions that are involved in the test through the quantile $z_{1-\alpha}$, one can wonder, by looking at Figure \ref{Z}, if it may perform reasonably well in practice with a usual number of a few tens of trials. 

Furthermore, looking informally at Equation \eqref{ConvUstat}, readers could think of two approximations that could be roughly  formulated in the following way:
\begin{equation}\label{approx1}
{\bf U}(\mathbb{X}_n^\indep)\overset{\mathcal{L}}{\underset{n\to\infty}{\approx}} \mathcal{N}\left(0, n \hat{\sigma}^2(\mathbb{X}_n^\indep)\right),\end{equation}
and
\begin{equation}\label{approx2}
 {\bf C}(\mathbb{X}_n^\indep)\overset{\mathcal{L}}{\underset{n\to\infty}{\approx}} \mathcal{N}\left(\hat{\bf C}_0(\mathbb{X}_n^\indep), n \hat{\sigma}^2(\mathbb{X}_n^\indep)\right).
\end{equation}

\begin{figure}[h!]
\vspace{-1.5cm}
\hspace{-2.3cm}
\includegraphics[scale=0.69]{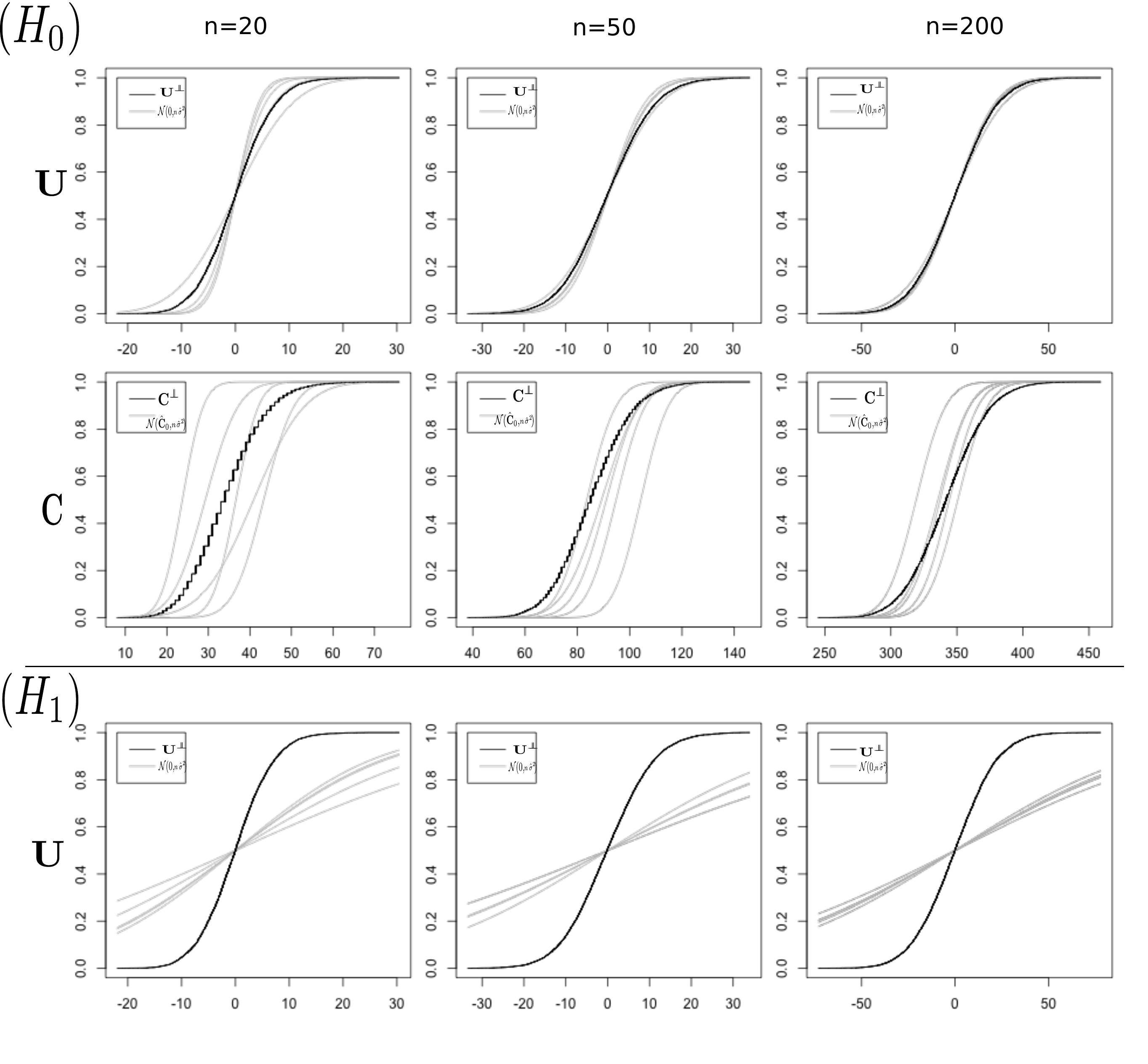}\\
\vspace{-2.2cm}
\caption{\label{Conv} Other Gaussian distribution approximations. Two first lines: c.d.f. of  ${\bf U}$ and ${\bf C}$ under $(H_0)$, obtained as in Figure \ref{Z}. These c.d.f. are respectively compared with the Gaussian c.d.f. with mean $0$ and standard deviation $\sqrt{n}\hat{\sigma}(\mathbb{X}_n)$, and the Gaussian c.d.f. with mean $\hat{\bf C}_0(\mathbb{X}_n)$ and standard deviation $\sqrt{n}\hat{\sigma}(\mathbb{X}_n)$, for five different simulations of $\mathbb{X}_n$ under $(H_0)$. Third line:  c.d.f. of ${\bf U}$ under $(H_0)$ computed as  above, compared with the centered Gaussian c.d.f. with standard deviation $\sqrt{n}\hat{\sigma}(\mathbb{X}_n)$, for five different simulations of $\mathbb{X}_n$ under $(H_1)$ (same marginals as in the first two lines but $X^1=X^2$). } 
\end{figure}

This is illustrated on Figure \ref{Conv}.

Looking at the first line of Figure \ref{Conv}, one can see that the approximation formulated in \eqref{approx1} is actually conceivable for large values of $n$. Note that in practice, one cannot have access to $\hat{\sigma}^2(\mathbb{X}_n^\indep)$ and it has to be replaced by $\hat{\sigma}^2(\mathbb{X}_n)$, meaning that it is computed with the observed sample. This does not change anything under $(H_0)$ since $\mathbb{X}_n$ is in this case distributed as $\mathbb{X}_n^\indep$. But this is a particularly important sticking point if $(H_0)$ is not satisfied as one can see on the third line of Figure \ref{Conv}:   the distribution of ${\bf U}(\mathbb{X}_n^\indep)$ does not look like  a centered Gaussian distribution of variance $n \hat{\sigma}^2(\mathbb{X}_n)$, when $\mathbb{X}_n$ does not satisfy $(H_0)$. 

More importantly, the second line of Figure \ref{Conv} shows that the approximation formulated in \eqref{approx2} is in fact misleading. To understand why, one needs to take into account the two following points.

(i) $\hat{\bf C}_0(\mathbb{X}_n^\indep)$ moves around its expectation $c_0$ (which is also the expectation of ${\bf C}(\mathbb{X}_n^\indep)$) with realizations of $\mathbb{X}_n^\indep$. These fluctuations have  an order of magnitude of $\sqrt{n}$ and are therefore perfectly observable on the  distribution of ${\bf C}(\mathbb{X}_n^\indep)$ whose variance is also of order $\sqrt{n}$.

(ii) $n\hat{\sigma}^2(\mathbb{X}_n^\indep)$ estimates the variance of ${\bf U}(\mathbb{X}_n^\indep)$ and not the one of ${\bf C}(\mathbb{X}_n^\indep)$ or $\hat{\bf C}_0(\mathbb{X}_n^\indep)$. This explains why not only the mean but also the variance  are  badly estimated in the second line of Figure \ref{Conv}. Both randomness (the one of ${\bf C}(\mathbb{X}_n^\indep)$ and the one of $\hat{\bf C}_0(\mathbb{X}_n^\indep)$) have to be taken into account to estimate the variance of ${\bf U}(\mathbb{X}_n^\indep)$.

As a conclusion of this first naive approach, the test of purely asymptotic nature, which consists in rejecting $(H_0)$ when ${\bf Z}^{obs}>z_{1-\alpha}$  may work  for $n$ large enough, as the variance is here computed by considering the correctly recentered statistic ${\bf U}$, and this even if the behavior of the statistic under $(H_1)$ is not clear. But an ad hoc and more naive test statistic, based on an estimation of the variance of ${\bf C}$ directly  and  without taking into account the fact that the centering term $\hat{\bf C}_0(\mathbb{X}_n)$ is also random,
would not  lead to a meaningful test.

\subsection{The bootstrap approaches}

It is well known \citep{ginesaintflour} that tests of purely asymptotic nature as the one presented above are less accurate for small values of $n$ than more involved procedures. In this article, the focus is on bootstrap/resampling procedures that are usually known to improve the performance from moderate to large sample sizes. Three main procedures are investigated: the trial-shuffling introduced in \citep{Pipa2003,Pipaet2003}, the classical full bootstrap of independence and the permutation approach \citep{romano}.

\begin{figure}[h!]
\begin{center}
\includegraphics[scale=0.5]{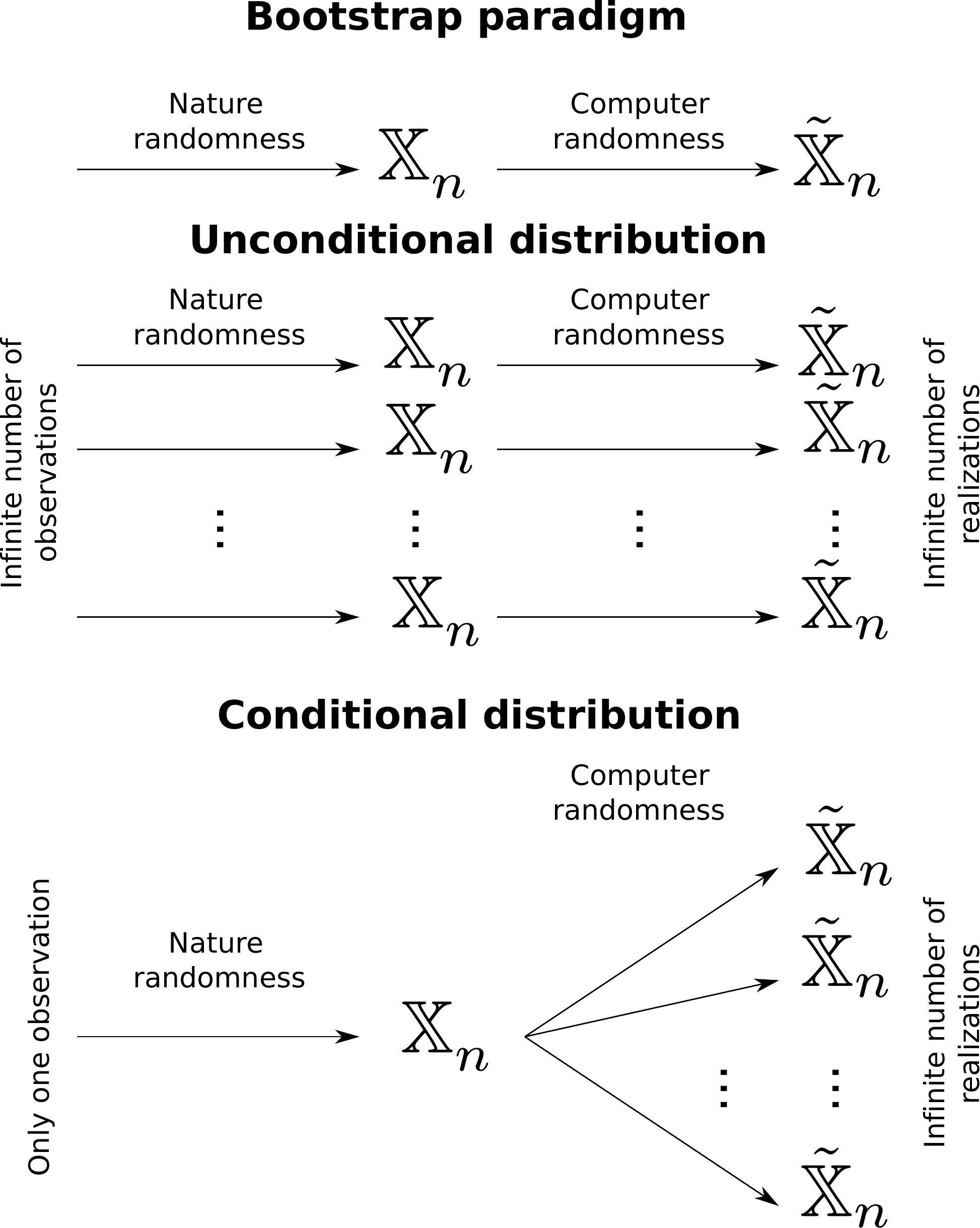}
\end{center}
\caption{\label{schema} Schematic view of the bootstrap paradigm and the difference between unconditional distribution and conditional distribution given the observation.}
\end{figure}

The main common paradigm of these three methods, as described in the sequel, is that  starting from an observation of the sample $\mathbb{X}_n$, they randomly generate via a computer another sample  $\tilde{\mathbb{X}}_n$, whose distribution should be close to the distribution of $\mathbb{X}_n^\indep$ (see also Figure \ref{schema}).

\bigskip

\Ovalbox{\begin{minipage}{14cm}
\begin{center}
{\bf Trial-shuffling}
\end{center}
$$\tilde{\mathbb{X}}_n=\mathbb{X}_n^{TS}=((X_{i^{TS}(1)}^1,X_{j^{TS}(1)}^2),...,(X_{i^{TS}(n)}^1,X_{j^{TS}(n)}^2)),$$
where the $(i^{TS}(k),j^{TS}(k))$'s are $n$ i.i.d. couples drawn uniformly at random in $\{(i,j)~/~ i=1,...,n, j=1,...,n, i\not=j\}$.

In particular, the corresponding bootstrapped coincidence count is
$${\bf C}^{TS}={\bf C}(\mathbb{X}_n^{TS}):=\sum_{k=1}^n \varphi\left(X_{i^{TS}(k)}^1,X_{j^{TS}(k)}^2\right).$$
\end{minipage}
}\\

This algorithm seems natural with respect to \eqref{indepegal} because it avoids the diagonal terms of the square $\{(i,j) ~/~ i=1,...,n,~j=1,...,n\}$. Hence as a result, $$\E({\bf C}^{TS})= c_0 = \E_{\indep}({\bf C}).$$

\bigskip

\Ovalbox{\begin{minipage}{14cm}
\begin{center}
{\bf Classical full bootstrap}
\end{center}
$$\tilde{\mathbb{X}}_n= \mathbb{X}_n^{*}=((X_{i^*(1)}^1,X_{j^*(1)}^2),...,(X_{i^*(n)}^1,X_{j^*(n)}^2)),$$
 where the $n$ couples $(i^*(k),j^*(k))$ are i.i.d. and where $i^*(k)$ and $j^*(k)$ are drawn uniformly and \underline{independently} at random in $\{1,...,n\}$. 

In particular, the corresponding bootstrapped coincidence count is
$${\bf C}^{*}={\bf C}(\mathbb{X}_n^{*}):=\sum_{k=1}^n \varphi(X_{i^*(k)}^1,X_{j^*(k)}^2).$$
 \end{minipage}
}\\

Note that this algorithm  draws uniformly at random in the square $\{(i,j) ~/~ i=1,...,n,~j=1,...,n\}$ and therefore does not avoid the diagonal terms. The idea behind this algorithm is to mimic the independence under $(H_0)$ of $X_k^1$ and $X_k^2$ by drawing the indexes $i^*(k)$ and $j^*(k)$ independently. However
$$ \E({\bf C}^*)= n \left[\frac{1}{n} \E(\varphi(X^1,X^2) )+\frac{n-1}{n} \E_\indep(\varphi(X^{1,\indep},X^{2,\indep}) ) \right].$$
Hence under $(H_0)$, $ \E_\indep({\bf C}^*)=c_0$ but, under $(H_1)$, $ \E({\bf C}^*)$ and $c_0$ are only asymptotically equivalent.

\bigskip
 
\Ovalbox{\begin{minipage}{14cm}
\begin{center}
{\bf Permutation}
\end{center} 
 $$\tilde{\mathbb{X}}_n= \mathbb{X}_n^{\Pi_n}=((X_{1}^1,X_{\Pi_n(1)}^2),...,(X_{n}^1,X_{\Pi_n(n)}^2)),$$
where $\Pi_n$ is a permutation drawn uniformly at random in the group of permutations $\mathfrak{S}_n$ of the set of indexes $\{1,\ldots,n\}$.

In particular, the corresponding bootstrapped coincidence count is
 $${\bf C}^\star={\bf C}\left(\mathbb{X}_n^{\Pi_n}\right):=\sum_{i=1}^n \varphi\left(X_i^1,X_{\Pi_n(i)}^2\right).$$
\end{minipage}
}\\

The idea is to use permutations to avoid  picking twice the same spike train of the same trial. In particular under $(H_0)$, the sum in ${\bf C}^\star$ is still a sum of independent variables, which is not the case in both of the previous algorithms. However, under $(H_1)$, the behavior is not as limpid. As for the full bootstrap, $$ \E({\bf C}^\star)= n \left[\frac{1}{n} \E(\varphi(X^1,X^2) )+\frac{n-1}{n} \E_\indep(\varphi(X^1,X^2) )\right] .$$
Hence under $(H_1)$, $ \E({\bf C}^*)$ and $c_0$ are only asymptotically equivalent.

\bigskip

To compare those three bootstrap/resampling algorithms, the first thing to wonder is whether, at least under $(H_0)$,  the introduced extra randomness has not impacted the distribution. More precisely, as stated above, all three procedures satisfy
$$\E_\indep({\bf C}(\tilde{\mathbb{X}}_n)) = \E_\indep({\bf C}(\mathbb{X}_n^\indep))=c_0,$$
but is the full distribution of ${\bf C}(\tilde{\mathbb{X}}_n)$ the same as the one of ${\bf C}(\mathbb{X}_n^\indep)$?

\begin{figure}
\hspace{-1.5cm}\includegraphics[scale=0.7]{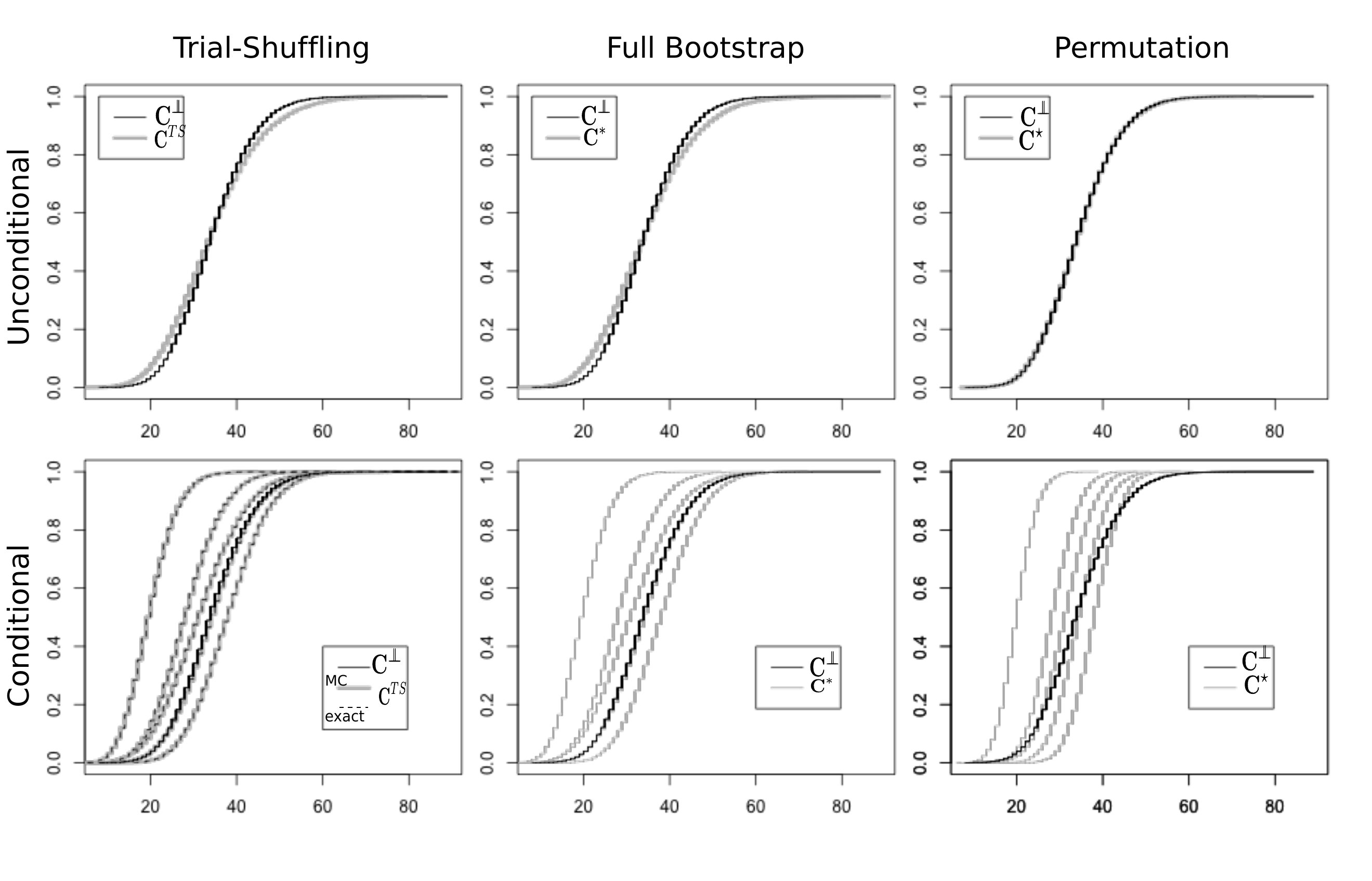}
\caption{\label{Uncond-cond} The unconditional distribution and conditional distributions of ${\bf C}$ under $(H_0)$.  C.d.f. of ${\bf C}(\mathbb{X}_n^\indep)$ and (for the first line) of ${\bf C}^{TS}= {\bf C}(\mathbb{X}_n^{TS})$, of ${\bf C}^{*}= {\bf C}(\mathbb{X}_n^{*})$  and of ${\bf C}^{\star}= {\bf C}(\mathbb{X}_n^{\Pi_n})$ obtained from $10000$ simulations of $n=20$ trials of two independent Poisson processes of firing rate $30$Hz on a window of length $0.1$s with $\delta=0.01$s. On the second line, in addition to the c.d.f. of ${\bf C}(\mathbb{X}_n^\indep)$, five observations of $\mathbb{X}_n=\mathbb{X}_n^\indep$ have been simulated in the same set-up and given these observations, the conditional c.d.f. have been approximated by simulating $10000$ times the extra-randomness corresponding to $\tilde{\mathbb{X}}_n$. For the trial-shuffling, in addition to this approximate Monte-Carlo method (MC), the exact conditional c.d.f. has been obtained thanks to the algorithm of \citep{Pipaet2003}.  }
\end{figure}


The first line of Figure \ref{Uncond-cond} shows as expected that the permutation does not change the distribution of $\mathbb{X}_n^\indep$, since, as said above, no spike train is picked twice. However, clearly the trial-shuffling and the full bootstrap have not the same property, even if the distributions are quite close.

Nevertheless, this is not completely convincing. Indeed, the main advantage of bootstrap procedures is to be able for one current observation of $\mathbb{X}_n$ to generate several  realizations of $\tilde{\mathbb{X}}_n$ to obtain not the unconditional distribution of ${\bf C}(\tilde{\mathbb{X}}_n)$ but the conditional distribution of ${\bf C}(\tilde{\mathbb{X}}_n)$ given $\mathbb{X}_n$. Figure \ref{schema} gives a more visual representation of the difference between conditional and unconditional distributions.

In particular, in a bootstrap testing procedure, the critical values are quantiles \footnote{In fact,  the quantiles are usually approximated by a Monte-Carlo method, since one has access to only a huge but still finite number of realizations of $\tilde{\mathbb{X}}_n$ given $\mathbb{X}_n$ in practice.}  of the conditional distribution of the bootstrapped test statistic given the observation of $\mathbb{X}_n$ and not the quantiles of the unconditional distribution. Hence, to see whether bootstrapped critical values associated to ${\bf C}(\mathbb{X}_n)$ are reasonable,   the conditional distribution of ${\bf C}(\tilde{\mathbb{X}}_n)$ given $\mathbb{X}_n$ has to be compared with the distribution of ${\bf C}(\mathbb{X}_n^\indep)$, and this whether  $\mathbb{X}_n$ satisfies $(H_0)$ or not.

The second line of Figure \ref{Uncond-cond} shows what happens for the approximated conditional distribution of ${\bf C}(\tilde{\mathbb{X}}_n)$ given $\mathbb{X}_n$ under $(H_0)$ in the three considered bootstrap approaches. Surprisingly none of these three conditional distributions seems to fit the distribution of ${\bf C}(\mathbb{X}_n^\indep)$. One may eventually think that this is due to the Monte-Carlo approximation of the conditional distributions, but for the trial-shuffling approach, Pipa and Gr\"un developed an algorithm for exact computation of the conditional distribution \citep{Pipaet2003}: both Monte-Carlo and exact conditional distribution are so close that it is difficult to make any difference between them. Hence there should be another explanation. In fact, the curves on the second line of Figure \ref{Uncond-cond} are similar to the ones on the second line of Figure \ref{Conv}. In both set-ups, one wonders if the distribution of ${\bf C}(\mathbb{X}_n^\indep)$ can or cannot be approximated by a distribution depending on the observation of $\mathbb{X}_n$:  a very basic Gaussian distribution for Figure \ref{Conv} and a more intricate distribution using the bootstrap paradigm for Figure ~\ref{Uncond-cond}. Nevertheless both distributions are too widely spread around the aim which is the distribution of ${\bf C}(\mathbb{X}_n^\indep)$. Since the explanation for Figure~\ref{Conv} was a centering defect that can be corrected by considering ${\bf U}$,  one of the possible explanation here is a centering defect for the bootstrap procedures too. 

\subsection{Which centering for which bootstrap ?}\label{secCent}

All the bootstrap approaches that have been proved to work from a mathematical point of view are based on centered quantities \citep{ginesaintflour}. In particular, the precursor work of Bickel and Freedman \citep{bickelfreedman} on the bootstrap of the mean can be heuristically explained as follows.

 Given a $n$ sample of i.i.d. real random variables $\mathbb{Y}_n=(Y_1,...,Y_n)$ with mean $m$ and  a corresponding bootstrap sample $\mathbb{Y}_n^*$, it is not possible to estimate the distribution of the empirical mean $\bar{Y}=(1/n) \sum_{i=1}^nY_i$ directly. However one can estimate the centered distribution, i.e. the distribution of $\bar{Y}-m=\bar{Y}-\E(\bar{Y})$. To do so, it is sufficient to replace "empirical mean" by "empirical bootstrap mean" and "expectation" by "conditional expectation". More explicitly, denoting by $\bar{Y}^*$ the empirical mean of the bootstrap sample $\mathbb{Y}_n^*$, the distribution of 
 $\bar{Y}-\E(\bar{Y})$
 is approximated by the conditional distribution given $\mathbb{Y}_n$ of
 $\bar{Y}^*-\E(\bar{Y}^*|\mathbb{Y}_n).$ 
 
 Transposed in our framework, this paradigm would mean that one can obtain a good fit of the distribution of $(1/n)({\bf C}(\mathbb{X}_n^\indep)-c_0)$ by the conditional distribution of $(1/n)({\bf C}(\tilde{\mathbb{X}}_n)-\E({\bf C}(\tilde{\mathbb{X}}_n)|\mathbb{X}_n))$ given $\mathbb{X}_n$. But as seen above, constructing a test with test statistic $(1/n)({\bf C}(\mathbb{X}_n)-c_0)$  is impossible in a full distribution free context where the value of $c_0$ is unknown.
 
 Therefore one needs to find a quantity that is completely observable but whose mean is still null under $(H_0)$. The statistic ${\bf U}$ introduced in Section \ref{naiv}  is suitable from this point of view. What one needs to check is whether the distribution of ${\bf U}(\mathbb{X}_n)$ under $(H_0)$, that is of   ${\bf U}(\mathbb{X}_n^\indep)$ (which has zero mean), is well approximated by the distribution of ${\bf U}\left(\tilde{\mathbb{X}}_n\right)-\E\left({\bf U}\left(\tilde{\mathbb{X}}_n\right)|\mathbb{X}_n\right)$.
 
 For the trial-shuffling, since 
$$
{\bf U}\left(\mathbb{X}_n^{TS}\right)= \sum_{k=1}^n \varphi\left(X^1_{i^{TS}(k)},  X^2_{j^{TS}(k)}\right) - \frac{1}{n-1} \sum_{k\not = k'} \varphi\left(X^1_{i^{TS}(k)},  X^2_{j^{TS}(k')}\right),$$
one can easily see that because the couple $\left(i^{TS}(k),  j^{TS}(k)\right)$ is drawn uniformly at random in the set of the $(i,j)$'s such that $i\not = j$ (set of cardinality $n(n-1)$),
\begin{eqnarray*}
\E\left({\bf U}\left(\mathbb{X}_n^{TS}\right)|\mathbb{X}_n\right)&=&\frac{1}{n-1} \sum_{i\not= j} \varphi\left(X_i^1,X_j^2\right)-\frac{1}{n}\sum_{i,j} \varphi\left(X_i^1,X_j^2\right)\\&=& \frac{ \hat{\bf C}_0\left(\mathbb{X}_n\right)-{\bf C}\left(\mathbb{X}_n\right)}{n}\\
&=&-\frac{{\bf U}\left(\mathbb{X}_n\right)}{n}.
\end{eqnarray*}
Hence the correct bootstrap statistic is 
$$\tilde{\bf U}^{TS}=\tilde{\bf U}\left(\mathbb{X}_n^{TS}\right)={\bf U}\left(\mathbb{X}_n^{TS}\right)+\frac{{\bf U}\left(\mathbb{X}_n\right)}{n}.$$
However similar computations show that the full bootstrap and the permutation satisfy
$$\E\left({\bf U}\left(\mathbb{X}_n^{*}\right)|\mathbb{X}_n\right)=\E\left({\bf U}\left(\mathbb{X}_n^{\Pi_n}\right)|\mathbb{X}_n\right)=0,$$
so ${\bf U}\left(\mathbb{X}_n^{*}\right)$ and ${\bf U}\left(\mathbb{X}_n^{\Pi_n}\right)$ can be used directly.

\begin{figure}[h!]
\vspace{-2cm}
\hspace{-3.1cm}\includegraphics[scale=0.77]{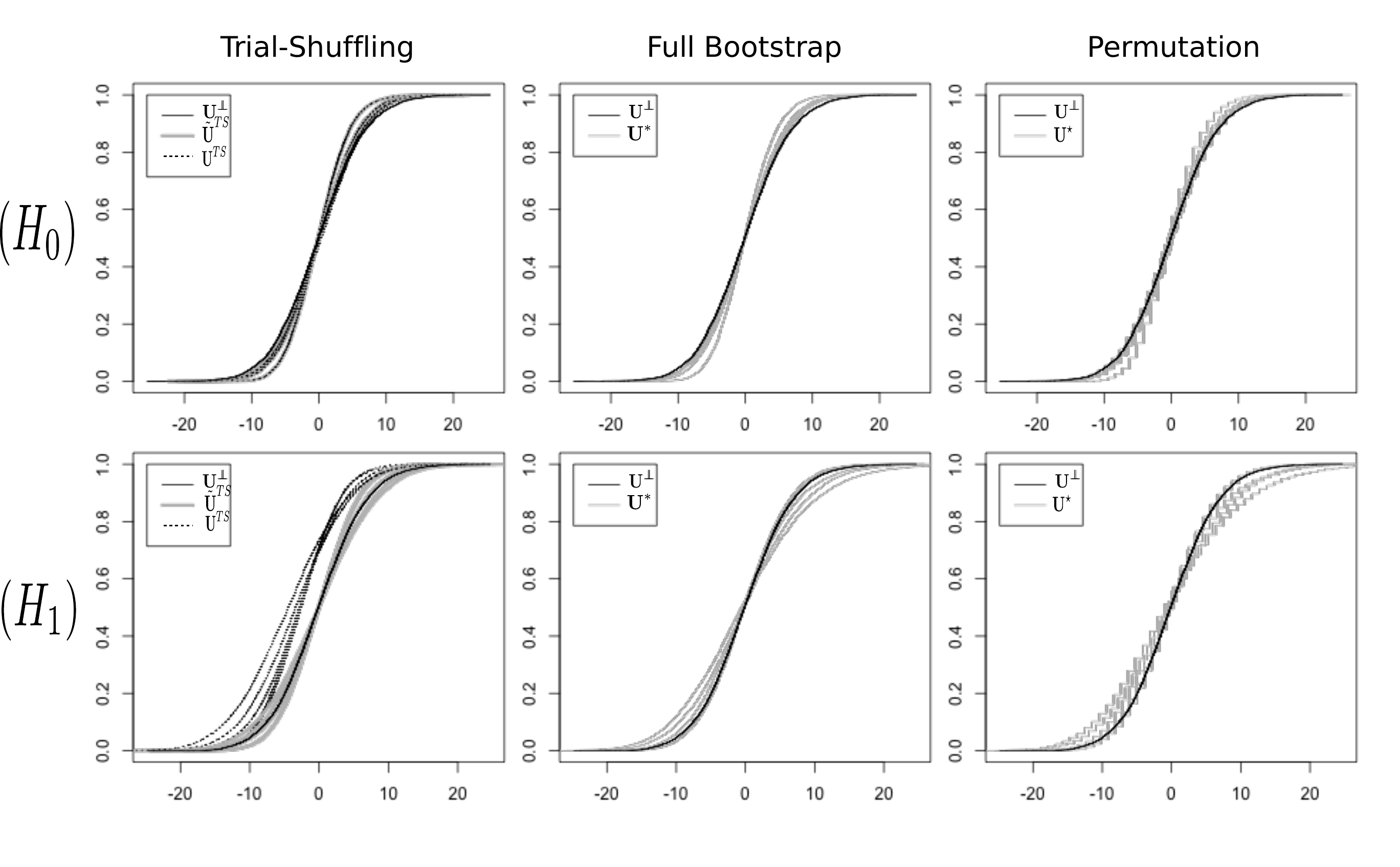}
\caption{\label{Condcentre} Conditional distribution of  ${\bf U}\left(\tilde{\mathbb{X}}_n\right)$ (or its recentered  version $\tilde{\bf U}^{TS}$ for the Trial-Shuffling) given $\mathbb{X}_n$. Cumulative distribution functions of ${\bf U}^\indep={\bf U}\left(\mathbb{X}_n^\indep\right)$ in black, obtained by simulation as in Figure  \ref{Uncond-cond}. For the first line, under $(H_0)$, five observations of $\mathbb{X}_n^\indep$  in the same set-up have been fixed and given these observations, the conditional c.d.f. of ${\bf U}^{TS}= {\bf U}\left(\mathbb{X}_n^{TS}\right)$, of $\tilde{\bf U}^{TS}={\bf U}^{TS}+{\bf U}^{obs}/n$, of ${\bf U}^{*}= {\bf U}\left(\mathbb{X}_n^{*}\right)$ and of ${\bf U}^{\star}= {\bf U}\left(\mathbb{X}_n^{\Pi_n}\right)$  have been obtained as in Figure  \ref{Uncond-cond}. For the second line,  five observations of $\mathbb{X}_n$, simulated under $(H_1)$ with marginals equal to the ones of the first line but satisfying $X^1=X^2$,  have been simulated and conditional c.d.f. are obtained in the same way as above.}\end{figure}

Figure \ref{Condcentre}  shows the quality of approximation of the distribution of ${\bf U}\left(\mathbb{X}_n^\indep\right)$ by the conditional distribution given the observation of either ${\bf U}^{*}= {\bf U}\left(\mathbb{X}_n^{*}\right)$ or  ${\bf U}^{\star}= {\bf U}\left(\mathbb{X}_n^{\Pi_n}\right)$. The approximation is accurate under $(H_0)$ but it is actually  also  accurate even if the observed sample is simulated under $(H_1)$, which is in complete accordance with the mathematical results of consistence in Wasserstein distance proved in \citep{bootnous}.   The approximation is just as accurate for the recentered statistic  $\tilde{\bf U}^{TS}={\bf U}^{TS}+{\bf U}^{obs}/n$.  Note that the difference between the  conditional c.d.f. of $\tilde{\bf U}^{TS}$ and the one of ${\bf U}^{TS}$ is particularly visible under $(H_1)$ when $X^1=X^2$. Hence, as explained by the computations above, in a trial-shuffling approach,  the recentered version leads to the correct bootstrap distribution. Note finally that this corroborates the  previous intuition: the reason why the approximation works for ${\bf U}$ and not for ${\bf C}$ is exactly the same as for the naive  approach of Figure \ref{Conv}. The centering is indeed random (here it can be viewed as $\E({\bf C}(\tilde{\mathbb{X}}_n)|\mathbb{X}_n)$) and one needs to take it into account  to have a correct approximation.

Finally an extra simplification holds in the permutation case, which may seem very surprising.
One can easily rewrite on the one hand,
$${\bf U}\left(\mathbb{X}_n\right)=\left(1-\frac{1}{n-1}\right) {\bf C}\left(\mathbb{X}_n\right)-\frac{1}{n-1} \sum_{i,j} \varphi\left(X_i^1,X_j^2\right)$$
and, on the other  hand, for the permutation sample
$${\bf U}\left(\mathbb{X}_n^{\Pi_n}\right)= \left(1-\frac{1}{n-1}\right) {\bf C}\left(\mathbb{X}_n^{\Pi_n}\right)-\frac{1}{n-1} \sum_{i,j} \varphi\left(X_i^1,X_j^2\right).$$
Note that the sum $\sum_{i,j} \varphi\left(X_i^1,X_j^2\right)$ is invariant by the action of the permutation. Hence if $u_t^\star$ denotes the quantile of order $t$ of the conditional distribution of  ${\bf U}\left(\mathbb{X}_n^{\Pi_n}\right)$ given $\mathbb{X}_n$ and if $c_t^\star$ denotes the quantile of order $t$ of the conditional distribution of  ${\bf C}\left(\mathbb{X}_n^{\Pi_n}\right)$ given $\mathbb{X}_n$, this very simple relationship holds
$$ u_t^\star = \left(1-\frac{1}{n-1}\right) c_t^\star-\frac{1}{n-1} \sum_{i,j} \varphi\left(X_i^1,X_j^2\right).$$
Hence the  test that rejects $(H_0)$ when ${\bf U}\left(\mathbb{X}_n\right)>u_{1-\alpha}^\star$ is exactly the one that rejects $(H_0)$ when
 ${\bf C}\left(\mathbb{X}_n\right)>c_{1-\alpha}^\star$. Therefore despite the fact that the conditional distribution of ${\bf C}\left(\mathbb{X}_n^{\Pi_n}\right)$ is not close at all to the one of ${\bf C}\left(\mathbb{X}_n^\indep\right)$, the test based on  ${\bf C}$ works, because it is equivalent to the test based on ${\bf U}$, for which the approximation of the conditional distribution works. 
Note however that this phenomenon happens only in the permutation approach, but not in the trial-shuffling or the full bootstrap approaches.

\subsection{Practical testing procedures and $p$-values}
From the considerations given above, five different tests  may be investigated, the first one based on a purely asymptotic approach, and the four other ones based on bootstrap approaches, with critical values approximated through a Monte-Carlo method. For each test, the corresponding $p$-values (i.e. the values of $\alpha$ for which the test passes from acceptance to rejection) are given. 

\paragraph{The naive test (N)}
It consists in rejecting  $(H_0)$ when $${\bf Z}^{obs}\geq  z_{1-\alpha}.$$
The corresponding $p$-value is given by: 
$$1-\Phi\left({\bf Z}^{obs}\right),$$
where $\Phi$ is the c.d.f. of a standard Gaussian distribution.

\paragraph{The Trial-Shuffling  test, version {\bf C} (TSC)}
It consists in rejecting $(H_0)$ when  
$${\bf C}^{obs} \geq \hat{c}_{1-\alpha}^{TS},$$
where $\hat{c}_{1-\alpha}^{TS}$ is the empirical quantile of order $(1-\alpha)$   of the conditional distribution of ${\bf C}^{TS}$ given $\mathbb{X}_n$. This empirical quantile is estimated over $B$ ($B=10000$ usually) realizations ${\bf C}^{TS}_1,...,{\bf C}^{TS}_B$ given the observed sample $\mathbb{X}_n$. The corresponding $p$-value is given by: 
$$\frac{1}{B} \sum_{i=1}^B {\bf 1}_{{\bf C}^{TS}_i\geq {\bf C}^{obs} }.$$
Despite the problems underlined in Section \ref{secCent}, we kept this test in the present study since it corresponds to the one programmed in \citep{Pipa2003} and since this bootstrap procedure is usually the one applied by neuroscientists.

\paragraph{The Trial-Shuffling  test, version recentered {\bf U} (TSU)}
It consists in rejecting $(H_0)$ when  
$${\bf U}^{obs} \geq \hat{w}_{1-\alpha}^{TS},$$
where $\hat{w}_{1-\alpha}^{TS}$ is the empirical quantile of order $(1-\alpha)$   of the conditional distribution of $\tilde{\bf U}^{TS}$ (the correctly recentered statistic)  given $\mathbb{X}_n$. This empirical quantile  and the corresponding $p$-value are  obtained in a similar way as for the above (TSC).

\paragraph{The Full Bootstrap test, version {\bf U} (FBU)}
It consists in rejecting $(H_0)$ when  
$${\bf U}^{obs}\geq \hat{u}_{1-\alpha}^{*},$$
where $\hat{u}_{1-\alpha}^{*}$ is the empirical quantile of order $(1-\alpha)$   of the conditional distribution of ${\bf U}^{*}$ given $\mathbb{X}_n$. This empirical quantile  and the corresponding $p$-value are  obtained in a similar way as for the above (TSC). 

\paragraph{The permutation test (P)}
The reader may think that it should consist in rejecting $(H_0)$ when  
$${\bf C}^{obs} \geq \hat{c}_{1-\alpha}^{\star},$$
where $\hat{c}_{1-\alpha}^{\star}$ is the empirical quantile of order $(1-\alpha)$   of the conditional distribution of ${\bf C}^{\star}$ given $\mathbb{X}_n$.
But the test by permutation is in fact  directly defined by its $p$-value, which is slightly different here, equal to:
$$\frac{1}{B+1} \left( 1 + \sum_{i=1}^B {\bf 1}_{{\bf C}^{*}_i\geq {\bf C}^{obs} } \right).$$
The permutation test then consists in rejecting $(H_0)$ when this $p$-value is less than $\alpha$.
Indeed, such a permutation test, with such a slightly different version of $p$-value, has been proved   to be exactly of level $\alpha$, whatever $B$ \citep{RW05}.

Note however that such a slight correction does not work for full bootstrap or trial-shuffling approaches, where the tests are only guaranteed to be asymptotically of level $\alpha$.

\begin{figure}
\hspace{-1.5cm}\includegraphics[scale=0.7]{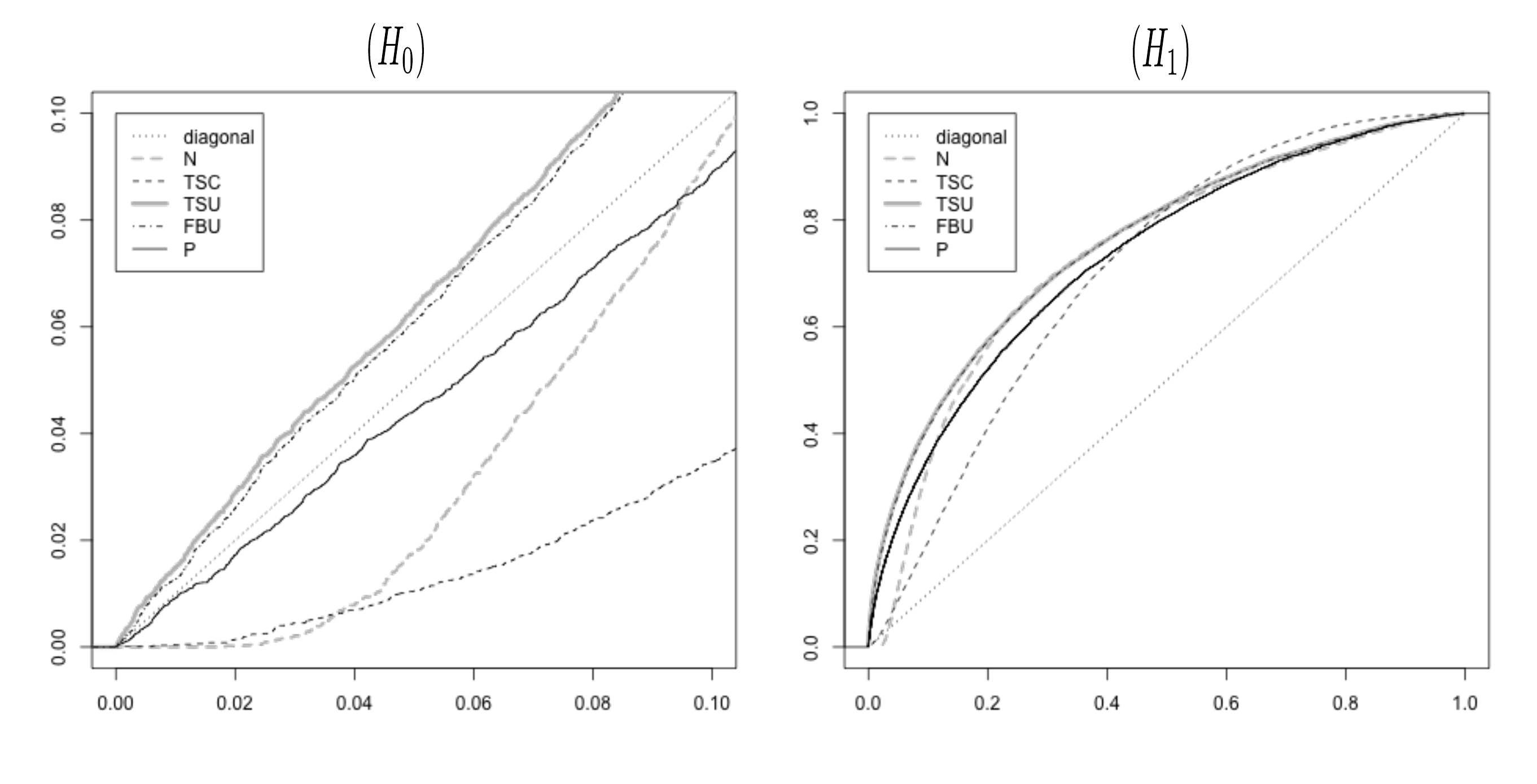}
\caption{\label{pval} Distribution of the $p$-values for the different tests. C.d.f. under both $(H_0)$ and $(H_1)$ of the $p$-values for the five tests: naive (N), Trial-Shuffling version {\bf C} (TSC), Trial-Shuffling version {\bf U} (TSU), Full Bootstrap version {\bf U} (FBU), and Permutation (P). Under $(H_0)$, the c.d.f. are obtained by simulations done as in Figure \ref{Uncond-cond}; the c.d.f. are then plotted only for small $p$-values ($\leq 0.1$). Under $(H_1)$, the couple $(X^1,X^2)$ is constructed by injection  \citep{GDA,mtgaue}, i.e. as $(N^1\cup N^{inj},N^2\cup N^{inj})$ where $(N^1,N^2)$ are two independent Poisson processes of firing rate $27$ Hz on a window of length $0.1$s and where $N^{inj}$ is a common Poisson process of firing rate $3$Hz; once again, $20$ i.i.d. trials are simulated $10000$ times to obtain the corresponding c.d.f. with $\delta=0.01$s.}
\end{figure}

 Saying that a test rejects at level $\alpha$ is exactly equivalent to saying that its $p$-value is less than $\alpha$. If a test is of level $\alpha$ for any $\alpha$ in $(0,1)$, the c.d.f. of its $p$-values should therefore be smaller that the one of a uniform variable (i.e. the diagonal). Between several tests with this guarantee, the less conservative one is the one for which the c.d.f of its $p$-values is the closest to the diagonal.
The left hand-side of Figure \ref{pval} shows the c.d.f. under $(H_0)$ of the corresponding $p$-values for the five considered  testing procedures and focuses on small $p$-values, which are the only  ones usually involved in testing,  to highlight the main differences between the five methods. For the chosen value of $n$ ($n=20$), the c.d.f. of the (TSU) and (FBU) $p$-values are almost identical and above the diagonal, meaning that the corresponding tests do not guarantee the level. On the contrary, the c.d.f. of the (N) and (TSC) $p$-values are clearly under the diagonal and far from it, meaning that the corresponding tests are too conservative. As guaranteed by \citep{RW05}, the permutation approach guarantees the level of the test: the c.d.f. of the (P) $p$-values is also under the diagonal, but much closer to the diagonal than the one of the (N) and (TSC) $p$-values. 

Furthermore, the behavior of the c.d.f. of the $p$-values under $(H_1)$ gives an indication of the power of the test:  the highest the c.d.f. of the $p$-values, the most powerful the corresponding test. Hence among the tests that guarantee the level, the permutation test (P) is the most powerful one. Note that other simulations in more various cases have been performed in \citep{bootnous} leading to the same conclusion. 

In the sequel, the focus is therefore on the permutation approach, keeping also the trial-shuffling version {\bf C} approach as a variant of the  method developed in \citep{Pipa2003}.

\section{Multiple tests}

\subsection{\label{complete}Description of the complete multiple testing algorithm}

To detect precise locations of dependence periods that can be matched to some experimental or behavioral events, it is classical to consider a family of windows $\mathcal{W}$ of cardinal $K$, which is a collection of potentially overlapping intervals $[a,b]$ covering the whole interval $[0,T]$ on which trials have been recorded \citep{grun,mtgaue}. Then, some independence tests are implemented on each window of the collection. Here we propose  a complete algorithm which takes into account  the multiplicity of the  tests, and which moreover enables to see if the coincidence count is significantly too large or too small on each window as in \citep{mtgaue}. \\

\hspace{-2cm}
\Ovalbox{
\begin{minipage}{17cm}
\begin{center}
{\bf Permutation UE algorithm} 
\end{center}

\medskip

Fix  a real number $q$ in $(0,0.5)$ and an integer $B$ larger than $2$.\\
\begin{tabular}{l}
- Do in parallel for all window $W=[a,b]$  in $\mathcal{W}$:\\
\begin{tabular}{ll}
&* Extract the points of the $X^1_i$'s  and $X_i^2$'s in $[a,b]$.\\
&* For all $(i,j)$ in $\{1,...,n\}^2$, compute $a_{i,j}=\varphi_\delta^{coinc}\left(X_i^1,X_{j}^2\right)$ over $[a,b]$ \\
& \hspace{0.3cm} by the {\bf delayed coincidence count algorithm}.\\
&* Draw at random $B$ i.i.d. permutations $\Pi_n^{\bf b}$, $1\leq {\bf b} \leq B$, and compute  ${\bf C}^{\bf b}=\sum_{i} a_{i,\Pi_n^{\bf b}(i)}$.\\
&* Compute also ${\bf C}^{obs}=\sum_{i} a_{i,i}.$ \\
&* Return 
$p^+_W=\frac{1}{B+1}\left(1+\sum_{{\bf b}=1}^B {\bf 1}_{{\bf C}^{\bf b}\geq {\bf C}^{obs}}\right)$ and
$p^-_W=\frac{1}{B+1}\left(1+\sum_{{\bf b}=1}^B {\bf 1}_{{\bf C}^{\bf b}\leq {\bf C}^{obs}}\right)$.
\end{tabular}\\
- Perform the procedure of \citep{BH} on the set of the above $2K$ $p$-values:\\
\begin{tabular}{ll}
&* Sort the $p$-values $p^{(1)}\leq ... \leq p^{(2K)}$.\\
&* Find $k = \max\{l ~/~ p^{(l)} \leq lq/(2K)\}$.\\
&* Return all the  $(W,\epsilon_W)$'s, for which $W$ is associated with one of the $p$-values $p^{(l)}$ for $l\leq k$, \\
& \hspace{0.3cm} with $\epsilon_W=1$ if $p^+_W\leq p^{(k)}$, so the coincidence count is significantly too large on $W$,\\
& \hspace{0.3cm} and $\epsilon_W=-1$ if $p^-_W\leq p^{(k)}$, so the coincidence count is significantly too small on $W$.\\
\end{tabular}

\end{tabular}
\end{minipage}}\\

This algorithm corresponds to a slight variation of the multiple testing step of \citep{mtgaue}, but adapted to non necessarily symmetric distributions \footnote{Note in particular that for a fixed $W$, one cannot have both $p^+_W<0.5$ and $p^-_W<0.5$ and therefore, if a $W$ is detected, it can only be because of one of the two  situations, $p^+_W\leq p^{(k)}$ or $p^-_W\leq p^{(k)}$, which cannot happen simultaneously.}. In several applications, neuroscientists are interested in detecting dependence periods for which the coincidence count is only significantly too large. In this case, one can use the restricted set of the $p^+_W$'s. Then if the considered windows are disjoint and if the spike trains are  Poisson processes that are non necessarily stationary, the False Discovery Rate (FDR) \footnote{see \citep{mtgaue}  or Table \ref{table} for a precise definition} of the above multiple testing procedure is  mathematically proved\footnote{The $p^+_W$'s are independent random variables such that $\mathbb{P}_{\indep}(p^+_W\leq \alpha)\leq \alpha$ for all $\alpha$ in $[0,1]$  \citep{BY,RW05}.} to be controlled by $q$ for any $B\geq 2$. 

 The code has been parallelized in \texttt{C++} and interfaced with \texttt{R}. The full corresponding  \texttt{R}-package is still a work in progress but actual codes are available at \\
 \texttt{https://github.com/ybouret/neuro-stat}.

 \subsection{Comparison on simulations}

\begin{figure}
\hspace{-3.2cm}\includegraphics[scale=0.5]{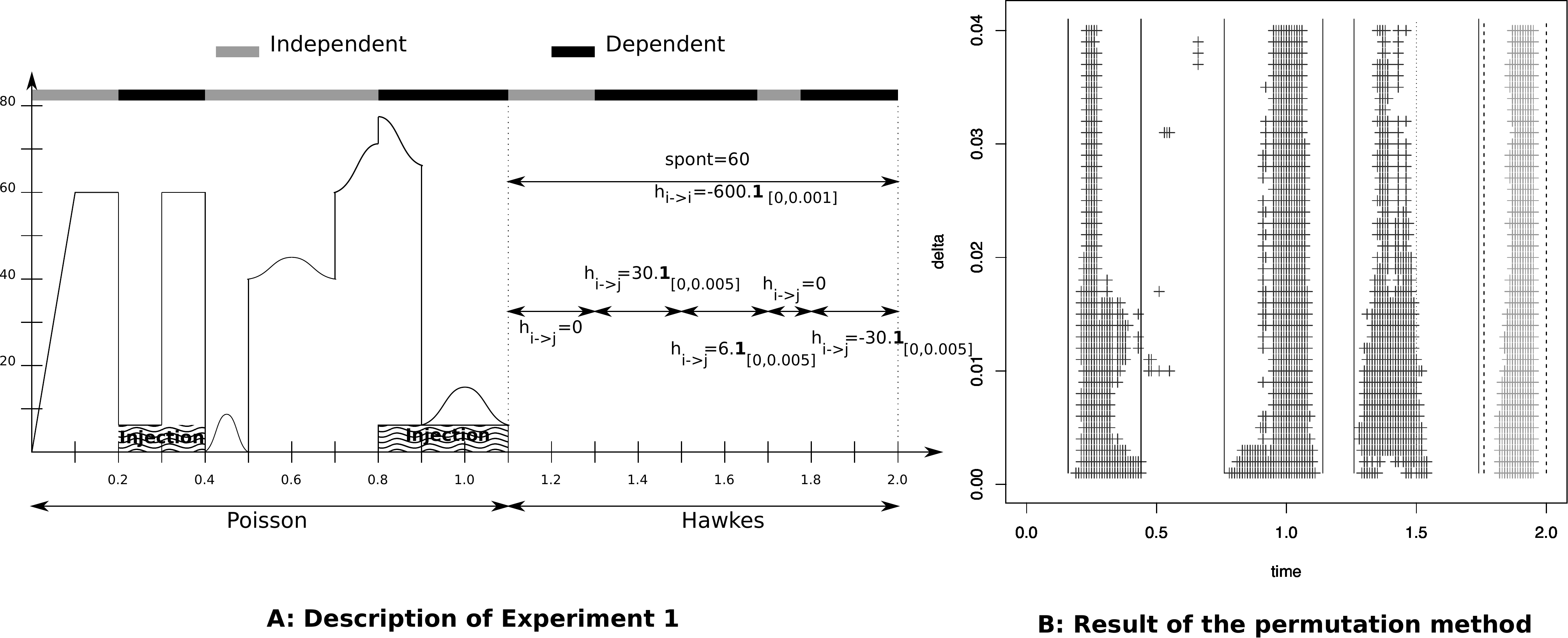}

\caption{\label{expfdr}Multiple tests. \ref{expfdr}.{\bf A}:  description of Experiment 1.  In the Poisson part, the intensity of both Poisson processes is plotted. The injection component corresponds to the part of a shared Poisson process which is injected in both processes corresponding to $X^1$ and $X^2$, as explained in Figure~\ref{pval}. In the Hawkes part (see \citep{mtgaue} for a complete description), formulas for the spontaneous parameters and both self interaction $h_{i\to i}$ and cross interaction $h_{i\to j}$ functions are given. \ref{expfdr}.{\bf B}: results of the Permutation UE method ($B=10000$, $q=0.05$) performed on $191$ overlapping windows of the form $[a,a+0.1]$ for $a$ in $\{0,0.01,...,1.9\}$ on one run of simulation for $50$ trials of Experiment~1. A black (resp. gray) cross is represented at the center of the window when it is detected by a $p^+_W$ (resp. $p^-_W$). Each horizontal line corresponds to a different $\delta$ in $\{0.001,0.002,...0.04\}$. The black vertical lines delimit the regions where the independence hypothesis is not satisfied: plain for positive dependence (i.e. where ${\bf C}^{obs}$ should be too large), and dashed for negative dependence (i.e. where ${\bf C}^{obs}$ should be too small). The dotted vertical line separates the region of high (on the left) and low (on the right) dependence in the Hawkes positive dependence part.}
\end{figure}

Two sets of simulations have been performed, the corresponding results are described in Table \ref{table} and one run of simulation of the Permutation UE  method is presented in Figure~\ref{expfdr}. Four methods have been compared: 
\begin{itemize}
\item the MTGAUE method of \cite{mtgaue} which assumes both processes to be homogeneous Poisson processes, 
\item the Trial-Shuffling, version {\bf C} (TSC) which corresponds to the method of \cite{Pipa2003}, which has been programmed with the delayed coincidence count described above and which has not been corrected for multiplicity.
\item the same as above but corrected by  Benjamini and Hochberg procedure (TSC + BH),
\item the Permutation UE approach described above. 
\end{itemize}
The permutation approach  always guarantees an FDR less than the prescribed level of $0.05$ whereas MTGAUE does not when the homogeneous Poisson assumption fails (Experiment 1). The classical trial-shuffling method (where dependence detection occurs each time the $p$-value is less than $0.05$)  seems to have comparable results in terms of both FDR and False Non Discovery Rate (FNDR) on Experiment 1  but fails to control the FDR on the most basic situation, namely purely independent processes (Experiment~2). Adding a  Benjamini-Hochberg step of selection of $p$-values to the trial-shuffling makes it more robust but at the price of a much larger FNDR with respect to the Permutation UE method, fact which is consistent with the conservativeness shown in Figure~\ref{pval}.

\begin{table}[h!]
\begin{minipage}[b]{5cm}
\begin{tabular}{|c|c|c|c|}
\hline
&Independ. & Depend.  & Total\\
\hline
Rejected & $V$ & $S$ & $R$\\
\hline
Accepted & $U$ & $T$ & $m-R$\\
\hline
Total & $m_0$ & $m-m_0$ & $m$\\
\hline
\end{tabular}
\end{minipage}
\hspace{3cm}\begin{minipage}[b]{6cm}
\begin{tabular}{|c|c|c|c|c|}
\hline
&\multicolumn{2}{|c|}{Experiment 1}&\multicolumn{2}{|c|}{Experiment 2} \\
& FDR  & FNDR & FDR & FNDR \\
\hline
MTGAUE & 0.10 & 0.17 & 0.04 & 0 \\
\hline
TSC & 0.01 & 0.26 & 0.25 & 0\\
\hline
TSC + BH & 0 & 0.32 & 0 & 0 \\
\hline
P & 0.01 & 0.23 & 0.02 & 0\\
\hline
\end{tabular}
\end{minipage}
\caption{\label{table} False Discovery and Non Discovery Rates. On the left hand-side, the classical table for multiple testing adapted to our dependence framework, with a total number of tests $m=2K$. On the right hand-side,  estimated FDR and FNDR over $1000$ runs, FDR being defined by $\E\left[(V/R){\bf 1}_{R>0}\right]$ and FNDR being defined by $\E\left[(T/(m-R)){\bf 1}_{m-R>0}\right]$. Experiment 1 is described in Figure \ref{expfdr}, Experiment 2 consists in two independent homogeneous Poisson processes of  firing rate $60$ Hz  on $[0,2]$. The set of windows is as in Figure \ref{expfdr}. There are $50$ trials and $\delta=0.01$s. MTGAUE is the method described in \citep{mtgaue} with $q=0.05$. (TSC) is the trial-shuffling method with Monte-Carlo approximation ($B=10000$) and the selected windows are the ones whose $p$-value are less than $0.05$. (TSC+BH) is the same method, except that the multiplicity of the tests is corrected by a Benjamini-Hochberg procedure ($q=0.05$). (P) corresponds to the Permutation UE method ($B=10000$, $q=0.05$).}
\end{table}
 
\subsection{Comparison on real data}

\paragraph{Behavioral procedure}
The data used in this theoretical article to test the dependence detection ability of the four methods were
already partially published in previous experimental studies \citep{RGDG,GR,Rieh2006} and also used in \citep{mtgaue}. These data were collected on a $5$-year-old male Rhesus monkey
who was trained to perform a delayed multidirectional pointing task. The animal sat in a
primate chair in front of a vertical panel on which seven touch-sensitive light-emitting diodes
were mounted, one in the center and six placed equidistantly ($60$ degrees apart) on a circle
around it. The monkey had to initiate a trial by touching and then holding with the left hand
the central target. After a fix delay of $500$ms, the preparatory signal (PS) was presented by
illuminating one of the six peripheral targets in green. After a delay of either $600$ms or $1200$ms,
selected at random with various probability, it turned red, serving as the response signal and
pointing target. During the first part of the delay, the probability $p_{resp}$ for the response signal to
occur at $(500+600)$ms $=1.1$s was $0.3$. Once this
moment passed without signal occurrence, the conditional probability for the signal to occur at $(500+600+600)$ms $=1.7$s changed to $1$. The monkey
was rewarded by a drop of juice after each correct trial. Reaction time (RT) was defined
as the release of the central target. Movement time (MT)
was defined as 
the touching of the
correct peripheral target.

\paragraph{Recording technique}
Signals recorded from up to seven microelectrodes (quartz insulated platinum-tungsten
electrodes, impedance: $2-5$M$\Omega$ at $1000$Hz) were amplified and band-pass filtered from
$300$Hz to $10$kHz. Using a window discriminator, spikes from only one single neuron per
electrode were then isolated. Neuronal data along with behavioral events (occurrences of
signals and performance of the animal) were stored on a PC for off-line analysis with a time
resolution of $10$kHz. 

\noindent
In the following study, only trials where the response signal (RS) occurs at $1.7$s are considered. The expected signal (ES) corresponds to an eventually expected but not confirmed signal, i.e. at $1.2$s. Only the pair $13$ of the previous data set is considered here, as it  was one of the main two examples already treated in \citep{mtgaue}. 

\begin{figure}
\hspace{-3.5cm}\includegraphics[scale=0.4]{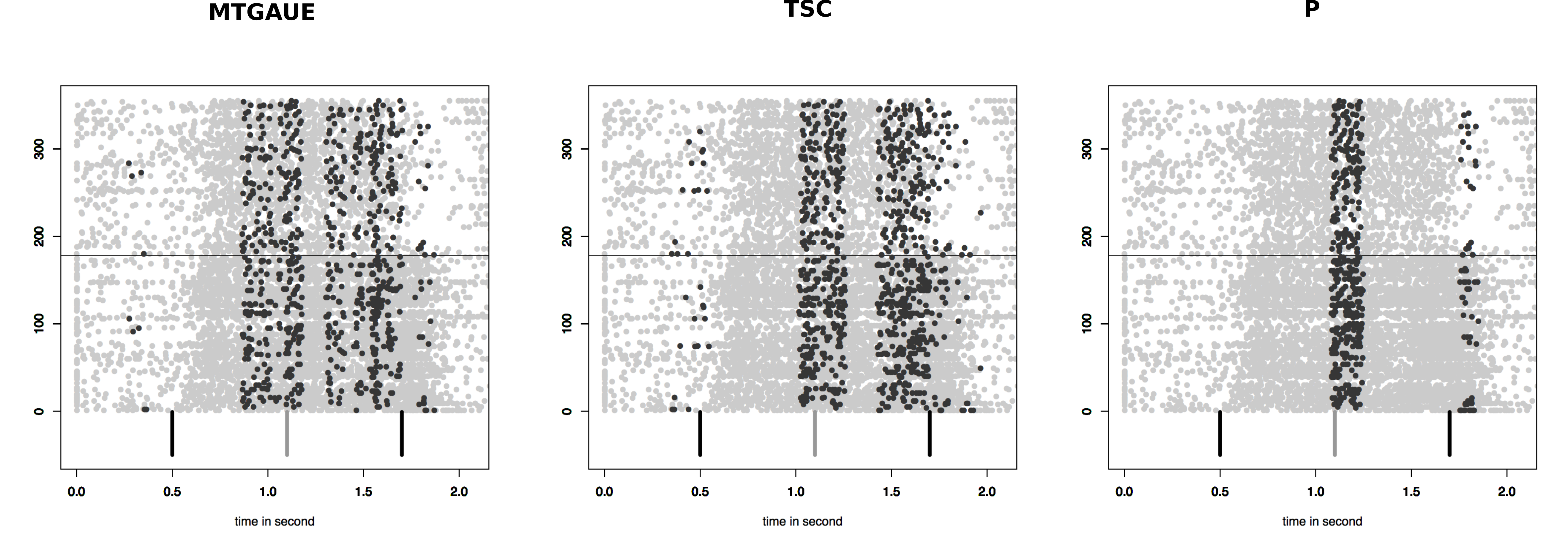}

\caption{\label{raster} Raster plots of the pair of neurons $13$. In black  the Unitary Events where the coincidence count is significantly too large  for the three methods (MTGAUE, TSC and P) presented in Table \ref{table}, with $\delta=0.02$s and $B=10000$. No interval was detected for a significantly too small coincidence count. Signs on bottom corresponds to behavioral events. The first black vertical bar corresponds to the preparatory signal (PS), the gray vertical bar to the expected signal (ES), the second black vertical bar to the response signal (RS).  }
\end{figure}

The results are presented in Figure \ref{raster}. The (TSC+BH) method does not detect anything and is therefore not presented. The Permutation UE method detects less windows than both (MTGAUE) and (TSC) methods, but the detected windows are still in adequation with the experimental or behavioral events. The above simulation study let us think that the extra detections of both (MTGAUE) and (TSC) may be false positives, since both methods do not control the FDR as well as the Permutation UE method.

\section{Discussion}

After describing a fast algorithm to compute the delayed coincidence count, showing that this notion can be used in practice for any surrogate data method in place of the binned coincidence count, we have focused on distribution free methods to test the independence between two simultaneously recorded spike trains. Though they are here presented with the delayed coincidence count, all these distribution free methods could be applied to any coincidence count if desired. 

Once the coincidence count ${\bf C}$ chosen, we have first introduced an empirical quantity or statistic ${\bf U}$ whose distribution is centered under the independence hypothesis $(H_0)$. In the spirit of \citep{mtgaue} but in a distribution free manner, a first naive method consists in performing a Gaussian approximation of the distribution of ${\bf U}$ under $(H_0)$. This method suffers from a not very sharp approximation when the number of trials $n$ is small (see Figure \ref{Z}). Moreover the approximation is clearly not valid when the observed sample does not satisfy $(H_0)$ (see the last line of Figure \ref{Conv}).

We then turned to bootstrap methods. One of the most well-known bootstrap method in the neuroscience literature is the trial-shuffling  \citep{Pipa2003,Pipaet2003}. It is usually based on a resampling approach directly applied on the coincidence count itself, namely ${\bf C}$. The other two investigated methods (full bootstrap and permutation) are well-known in the statistics literature for independence testing between real valued random vectors \citep{Hoeff,romano} and have been recently applied to point processes that are modeling here simultaneously recorded spike trains \citep{bootnous}. These last two methods are usually applied on centered statistics.

One of the main message of the present work is that applying bootstrap methods to non centered statistics, as ${\bf C}$, does not lead to a correct approximation of the distribution of ${\bf C}$ under  $(H_0)$ (see the second line of Figure \ref{Uncond-cond}). This phenomenon, combined with observations in the more classical framework of the naive test on Figure \ref{Conv}, leads us to think about a centering defect. Once the methods are applied on correct centered statistics ({\bf U} or $\tilde{\bf U}$ depending on the chosen bootstrap method), they all three outperform the naive approach. The approximation is better under $(H_0)$ for small value of $n$ (first line of  Figure \ref{Conv} and first line of Figure \ref{Condcentre}) and is still accurate when the observed sample does not satisfy $(H_0)$ (last line of  Figure \ref{Conv} and second line of Figure \ref{Condcentre}).

From an algorithmic point of view, all the corresponding bootstrap $p$-values are evaluated thanks to a Monte-Carlo algorithm and a program which interfaces \texttt{R} and \texttt{C++}, thus making the running time fast and the use easy. Pipa and Gr\"un \citep{Pipaet2003} have given an exact algorithm when the trial-shuffling is applied on the coincidence count {\bf C} directly. It is a very elegant algorithm using the fact that ${\bf C}$ is an integer that can take a small number of values. Unfortunately the same gain is not really possible for ${\bf U}$ which is not an integer and which can take much more values. Moreover this exact algorithm is quite long with respect to the  Monte-Carlo algorithm  when the number of simulations is $10000$ (as used in the present work) and one can see on the bottom left of Figure \ref{Uncond-cond} that the difference between both results (\ Monte-Carlo and exact algorithms) is not detectable at first glance. 

A more precise study of the Monte-Carlo approximated bootstrap  $p$-values shows that for a small number of trials, the trial-shuffling and the full bootstrap methods, even applied to a correctly centered statistic, do not provide tests of  prescribed level. On the contrary, the permutation method, thanks to an adequate version of its $p$-values \citep{RW05}, allows for a precise control of the level. The classical trial-shuffling method based on the non centered quantity {\bf C} and the naive approach also both lead to a precise control of the level but in a more conservative way (see the right-hand side of Figure \ref{pval}). This is also showed by the behavior of the $p$-values under the alternative, $p$-values that are smaller for the permutation approach than for the other two methods (see the right-hand side of Figure \ref{pval}).

Finally, we decided to combine the delayed coincidence count, which is much more precise than the binned coincidence count \citep{mtgaue,grun} with the permutation approach, and to apply the obtained independence testing procedure to several windows of detection simultaneously. The final proposed method consists in combining the individual tests with the  approach of \citep{BH} to correct for the multiplicity of the tests. Parallel programming is used to treat each window in an independent manner. This new algorithm named Permutation UE is completely distribution free. It better controls the False Discovery Rate than MTGAUE \citep{mtgaue} or the classical trial-shuffling method applied on {\bf C}  (see Table \ref{table}, methods (MTGAUE) and (TSC)).  
Moreover, it does not suffer from conservativeness as the trial-shuffling method applied on {\bf C}, once the multiplicity of the tests is taken into account (see Table \ref{table}, method (TSC+BH)). On real data, the results are similar to existing methods (MTGAUE, TSC) except for some detections that disappear but  that are likely to be false positive thanks to the present simulation study.

To conclude, we introduce in this article the Permutation UE method, which is a Unitary Events method based on delayed coincidence count and on an evaluation of $p$-values via a distribution free Monte-Carlo approximated permutation approach. This method does not suffer from any loss in synchrony detection as the binned coincidence count \citep{grunt}, is distribution free  and in this sense upgrades \citep{mtgaue}. Moreover the algorithm is fast and parallelized, and despite using a Monte-Carlo scheme, it can guarantee the single tests to be of the prescribed level and the multiple test to control the FDR in a non asymptotic manner, therefore outperforming the trial-shuffling method \citep{Pipa2003,Pipaet2003} in terms of both mathematical caution and computing time, when compared with the exact algorithm described in \citep{Pipaet2003}. Finally it is still sufficiently sensitive to detect reasonable features on real data sets. The only drawback is that it can only work for pairs of neurons. The definition of delayed coincidence count for more than two neurons has been recently introduced in \cite{julien_thomas}, but the combination of this notion with a bootstrap approach is still an open question.

\subsection*{Acknowledgments}
We  thank F. Grammont  for fruitful discussions. This work was granted access to the HPC and visualization resources of "Centre de Calcul Interactif" hosted by "Universit\'e Nice Sophia Antipolis". 
It was partly supported by the french Agence Nationale de la Recherche (ANR 2011 BS01 010 01 projet Calibration), by the PEPS BMI 2012-2013 {\it Estimation of dependence graphs for thalamo-cortical neurons and multivariate Hawkes processes} and by  the interdisciplanary axis MTC-NSC of the University of Nice Sophia-Antipolis. The PhD grant of M. Albert is funded by the PACA french region. 


\end{document}